\documentclass[showpacs,preprintnumbers,amsmath,amssymb]{revtex4}
\usepackage[dvips]{graphicx}

\begin{document}
\newcommand {\nc} {\newcommand}
\nc {\beq} {\begin{eqnarray}}
\nc {\eol} {\nonumber \\}
\nc {\eoln}[1] {\label {#1} \\}
\nc {\eeq} {\end{eqnarray}}
\nc {\eeqn}[1] {\label {#1} \end{eqnarray}}
\nc {\dem} {\mbox{$\frac{1}{2}$}}
\nc {\ve} [1] {\mbox{\boldmath $#1$}}
\renewcommand{\arraystretch}{1.0}

\title{Three-body continuum states on a Lagrange mesh}
\author{P. Descouvemont
}
\thanks{Directeur de Recherches FNRS}
\affiliation{Physique Nucl\'{e}aire Th\'{e}orique et Physique Math\'{e}matique, CP229\\
Universit\'{e} Libre de Bruxelles, B1050 Brussels, Belgium}
\author{E. M. Tursunov
}
\thanks{Permanent address: Institute of Nuclear Physics, 702132, Ulugbek,
Tashkent, Uzbekistan}
\affiliation{Physique Nucl\'{e}aire Th\'{e}orique et Physique Math\'{e}matique, CP229\\
Universit\'{e} Libre de Bruxelles, B1050 Brussels, Belgium}
\author{D. Baye}
\affiliation{Physique Nucl\'{e}aire Th\'{e}orique et Physique Math\'{e}matique, CP229\\
Universit\'{e} Libre de Bruxelles, B1050 Brussels, Belgium}
\affiliation{Physique Quantique, CP165/82, \\
Universit\'{e} Libre de Bruxelles, B-1050 Brussels, Belgium}
\begin{abstract}
Three-body continuum states are investigated with the hyperspherical method on a Lagrange mesh. The
$R$-matrix theory is used to treat the asymptotic behaviour of scattering wave functions. The formalism
is developed for neutral as well as for charged systems. We point out some specificities of continuum
states in the hyperspherical method. The collision matrix can be determined with a
good accuracy by
using propagation techniques. The method is applied to the $^6$He (=$\alpha$+n+n) and $^6$Be (=$\alpha$+p+p)
systems, as well as to $^{14}$Be (=$^{12}$Be+n+n). For $^6$He, we essentially recover results
of the literature. Application to $^{14}$Be suggests the existence of an excited $2^+$
state below threshold. The calculated B(E2) value should make this state observable with
Coulomb excitation experiments.
\end{abstract}
\maketitle
\section{Introduction}
Three-body systems present a large variety of interesting features \cite{Jo04,ZDF93}. The discovery of a halo structure in $^6$He \cite{THH85} triggered many experimental and theoretical works on exotic nuclei, such as $^6$He, $^{11}$Li or $^{14}$Be. The bound-state spectroscopy of Borromean systems is now relatively well known. On the experimental side, current intensities of radioactive beams are high enough for precise measurements of spectroscopic properties, such as energies, r.m.s. radii on quadrupole moments. On the theoretical side, several methods have been developed, and provide accurate solutions of the three-body Schr\"odinger equation.

The hyperspherical harmonic method (HHM) is known to be well adapted to three-body systems \cite{MF53,Li95}. The six Jacobi coordinates are replaced by five angles, and a single dimensional coordinate, the hyperradius. The HHM transforms the three-body Schr\"odinger equation into a set of coupled differential equations depending on the  hyperradius. It has been applied to many exotic nuclei.

Recently, we have combined the HHM with the Lagrange-mesh technique \cite{DDB03}. The Lagrange-mesh method (see Ref. \cite{BHV02} and references therein) is an approximate variational calculation that resembles a mesh calculation. The matrix elements are calculated at the Gauss approximation associated with
the mesh. They become very simple. In particular, the potential matrix elements are replaced by their values at the mesh points. In spite of its simplicity, the Lagrange-mesh method is as accurate as the corresponding variational calculation. This was shown for two-body \cite{BHV02} as well as for three-body \cite{DDB03} systems.

In the present work, we extend the formalism of Ref. \cite{DDB03} to three-body continuum states. The information provided by continuum states is a natural complement to the bound-state spectroscopy. Experimentally, three-body continuum states are investigated through breakup experiments (see for example Ref.~\cite{AAA99}). On the theoretical point of view, various methods have been developed. Some of them, such as the Complex Scaling Method \cite{Ho83}, or the Analytic Continuation in the Coupling Constant \cite{KK77} deal with resonances only; they cannot be applied to non-resonant states. Other methods, such as the $R$-matrix theory \cite{LT58} are more difficult to apply, but can be used for non-resonant, as well as for resonant states.

Applications of the $R$-matrix method to two-body systems have been performed for many years in nuclear as well as in atomic physics. In nuclear physics, applications to three-body systems are more recent \cite{TDE00}. The $R$-matrix theory allows the use of a variational basis to describe unbound states. It is based on an internal region, where the wave function is expanded over
the basis, and on an external region,
where the asymptotic behaviour can be used. In three-body systems, the hyperspherical formalism is very efficient for bound states. For unbound states, however, it raises problems owing to the long range of the coupling potentials \cite{TDE00}. In the $R$-matrix framework this can be solved by using propagation techniques \cite{BN95}.

In two-body systems, the Lagrange-mesh technique associated with the $R$-matrix formalism has
been applied in single- \cite{BHS98} and multi-channel \cite{HSV98} calculations. The purpose
of the present work is to extend the method to three-body systems. Another development
concerns the application to charged systems. Many exotic nuclei are unbound, even in
their ground states, due to the Coulomb force. We show applications to the $\alpha$+n+n
and $^{12}$Be+n+n systems, for which two-body potentials are available in the literature.
The mirror systems are also investigated.

In Section 2, we summarize the three-body formalism, and present the $R$-matrix method. Section 3 is devoted to applications to $^6$He and $^{14}$Be, with the mirror systems. Concluding remarks are given in Section 4.

\section{Three-body continuum states}
\subsection{Hamiltonian and wave functions}
Let us consider three particles with mass numbers $A_i$ (in units of the nucleon mass $m_N$), and space coordinates $\ve{r}_i$. A three-body Hamiltonian is given by
\beq
H=\sum_{i=1}^3T_i \ + \ \sum_{i>j=1}^3V_{ij}(\ve{r}_j-\ve{r}_i),
\label{eq1}
\eeq
where $T_i$ is the kinetic energy of nucleon $i$, and $V_{ij}$ a nucleus-nucleus potential. We neglect three-body forces in this presentation.

The HHM is known to be an efficient tool to deal with three-body systems. This formalism is well known, and we refer to Refs. \cite{ZDF93,Li95} for detail. Starting from coordinates $\ve{r}_i$, one defines the Jacobi coordinates $\ve{x}_k$ and $\ve{y}_k$ ($k=1,2,3$). We adopt here the notations of Ref. \cite{DDB03}. The hyperradius $\rho$ and hyperangle $\alpha_k$ are then defined as
\beq
& \rho^2 &= x_k^2+y_k^2, \nonumber \\
& \alpha_k &= \arctan{\frac{y_k}{x_k}}.
\label{eq2}
\eeq

The hyperangle $\alpha_k$ and the orientations $\Omega_{x_{k}}$ and $\Omega_{y_{k}}$ provide a set of angles $\Omega_{5k}$. In this notation the kinetic energy reads
\beq
T_{\rho}=\sum_{i=1}^3 T_i-T_{cm} = -\frac{\hbar^2}{2m_N} \left( \frac{\partial^2}{\partial \rho^2} +
\frac{5}{\rho} \frac{\partial}{\partial \rho} - \frac{K^2(\Omega_{5k})}{\rho^2}\right).
\label{eq3}
\eeq
In Eq.~(\ref{eq3}), $T_{cm}$ is the c.m. kinetic energy, and
$K^2$ is a five-dimensional angular momentum \cite{RR70} whose eigenfunctions (with eigenvalues $K(K+4)$) are given by
\beq
{\cal Y}^{\ell_x \ell_y}_{KLM_L}(\Omega_{5})&=&\phi^{\ell_x \ell_y}_{K}(\alpha)
\left[ Y_{\ell_x}(\Omega_{x})\otimes Y_{\ell_y}(\Omega_{y}) \right] ^{LM_L}, \nonumber \\
 \phi^{\ell_x \ell_y}_{K}(\alpha)&=&
{\cal N}_K^{\ell_x \ell_y} (\cos \alpha)^{\ell_x} (\sin \alpha)^{\ell_y}
P_n^{(\ell_y+\dem,\ell_x+\dem)}(\cos 2\alpha),
\label{eq4}
\eeq
where $P_n^{(\alpha,\beta)}(x)$ is a Jacobi polynomial and ${\cal N}_K^{\ell_x \ell_y}$ a normalization factor \cite{Li95} (here $k$ is implied).
In these definitions, $K$ is the hypermomentum, ($\ell_x,\ell_y$) are the orbital momenta
associated with ($\ve{x}$,$\ve{y}$), and $n$ is a positive integer defined by
\beq
n=(K-\ell_x-\ell_y)/2.
\label{eq4b}
\eeq
Introducing the spin component $\chi^{SM_S}$ yields the hyperspherical function with total spin $J$
\beq
{\cal Y}^{JM}_{\gamma K}(\Omega_5)=\left[ {\cal Y}^{\ell_x \ell_y}_{KL}(\Omega_5) \otimes \chi^{S}
\right] ^{JM},
\label{eq5}
\eeq
where index $\gamma$ stands for ($\ell_x,\ell_y,L,S$).

A wave function $\Psi^{JM\pi}$, solution of the Schr\"odinger equation associated with Hamiltonian (\ref{eq1}), is expanded over basis functions (\ref{eq5}) as
\beq
\Psi^{JM\pi}(\rho,\Omega_5)=\rho^{-5/2}\sum_{\gamma K} {\chi}^{J\pi}_{\gamma K}(\rho)\
{\cal Y}^{JM}_{\gamma K}(\Omega_5),
\label{eq6}
\eeq
where $\pi=(-1)^K$ is the parity of the three-body relative motion,
and ${\chi}^{J\pi}_{\gamma K}(\rho)$ are hyperradial wave functions which should be determined. Rigorously, the summation over ($\gamma K$) should contain an infinite number of terms. In practice, this expansion is limited by a maximum $K$ value, denoted as $K_{max}$. For weakly bound states, it is well known that the convergence is rather slow, and that large $K_{max}$ values must be used. Typically $100-200$ terms are necessary for realistic $K_{max}$ values.

The radial functions ${\chi}^{J\pi}_{\gamma K}(\rho)$ are derived from a set of coupled differential equations
\beq
&&\left[-\frac{\hbar^2}{2m_N} \left( \frac{d^2}{d\rho^2} - \frac{{\cal L}_K ({\cal L}_K + 1)}{\rho^2}\right) -E \right]
{\chi}^{J\pi}_{\gamma K}(\rho)\nonumber \\
&&+ \sum_{K' \gamma'} V^{J\pi}_{K  \gamma,K' \gamma'}(\rho)\, {\chi}^{J\pi}_{\gamma' K'}(\rho)=0,
\label{eq7}
\eeq
with ${\cal L}_K =K+3/2.$ The potential terms are given by the contribution of the three nucleus-nucleus interactions
\beq
V^{J\pi}_{K  \gamma,K' \gamma'}(\rho)=\sum_{i=1}^3
(V^{J\pi(Ni)}_{K  \gamma,K' \gamma'}(\rho)+V^{J\pi(Ci)}_{K  \gamma,K' \gamma'}(\rho)),
\label{eq8}
\eeq
where we have explicitly written the nuclear $(N)$ and Coulomb ($C)$ terms.

Assuming the use of ($\ve{x}_1,\ve{y}_1$) for the coordinate set, the contribution $i=1$ is directly determined from
\beq
V^{J\pi(1)}_{K  \gamma,K' \gamma'}(\rho)=\int
{\cal Y}^{JM*}_{\gamma K}(\Omega_5)\, V_{23}\left( \frac{\rho \cos \alpha}{\sqrt{\mu_{23}}} \right) \,
{\cal Y}^{JM}_{\gamma' K'}(\Omega_5) d\Omega_5,
\label{eq9}
\eeq
where $\mu_{ij}=A_iA_j/(A_i+A_j)$.
The terms $i=2,3$ are computed in the same way, with an additional transformation using the Raynal-Revai coefficients \cite{RR70}. Definition (\ref{eq9}) is common to the nuclear and Coulomb contributions. Integrations over $\Omega_x$ and $\Omega_y$ are performed analytically, whereas integration over the hyperangle $\alpha$ is treated numerically. For the Coulomb potential, the $\rho$ dependence is trivial; we have
\beq
\sum_{i=1}^3 V^{J\pi(Ci)}_{K  \gamma,K' \gamma'}(\rho)=z^{J\pi}_{K  \gamma,K' \gamma'}\frac{ e^2}{\rho}
\label{eq10}
\eeq
where $z^{J\pi}_{\gamma K,\gamma',K'}$ is an effective charge, independent of $\rho$, and calculated numerically from Eq.~(\ref{eq9}) and from Raynal-Revai coefficients \cite{VNA01}. Examples of matrices $z^{J\pi}$ are given in Ref.~\cite{VNA01}
for the $\alpha$+p+p system. Knowing the analytical $\rho$-dependence of the potential is crucial for continuum states (see below). Notice that, to derive Eq.~(\ref{eq10}), one assumes the
$1/|\ve{r}_j-\ve{r}_i|$ dependence of the Coulomb potential. Using a point-sphere definition is straightforward, as the difference
can be included in the nuclear part.

\subsection{Asymptotic behaviour of the potential}
For small $\rho$ values the potential must be determined by numerical integration of Eq. (\ref{eq9}). However, analytical approximations can be derived for large $\rho$ values. For the Coulomb interaction, definition (\ref{eq10}) is always valid. Let us now consider the nuclear contribution. After integration over $\Omega_x$ and $\Omega_y$, a matrix element between basis states (\ref{eq4}) is written as
\pagebreak
\beq
V^{\ell_x \ell_y,\ell'_x \ell'_y}_{KL,K'L'}(\rho)&=&
\delta_{LL'}\delta_{\ell_y \ell'_y}
\int_0^{\pi/2}
\phi^{\ell_x \ell_y}_{K}(\alpha)
V_{N}\left( \frac{\rho \cos \alpha}{\sqrt{\mu_{23}}} \right)
\phi^{\ell'_x \ell_y}_{K'}(\alpha)
\sin^2\alpha \cos^2\alpha d\alpha \nonumber \\
&=&{\cal N}_K^{\ell_x \ell_y}{\cal N}_{K'}^{\ell'_x \ell_y} \delta_{LL'}\delta_{\ell_y \ell'_y}
\frac{1}{\rho^3} \int_0^{\rho} P_n^{(\ell_y+\dem,\ell_x+\dem)}\left( 2\frac{u^2}{\rho^2}-1 \right)
V_N\left( \frac{u}{\sqrt{\mu_{23}}} \right) \nonumber \\
&& \times P_{n'}^{(\ell_y+\dem,\ell'_x+\dem)}\left( 2\frac{u^2}{\rho^2}-1 \right)
\left(1-\frac{u^2}{\rho^2}\right)^{\ell_y+\dem}
\left( \frac{u}{\rho} \right) ^{\ell_x+\ell'_x}u^2 du
\label{eq11}
\eeq

To deal with the spin, the coupling order in Eq.~(\ref{eq5}) is modified in order to introduce the
total spin of the interacting particles $\ve{j_x=\ell_x+S}$. This is achieved with standard
angular-momentum algebra, involving $6j$ coefficients. If the tensor force is not included, we also have $\ell_x=\ell'_x$.
For large $\rho$ values, and if the potential goes to zero faster than $1/u^2$, we can use the following expansions \cite{AS72}
\beq
&&P_n^{(\alpha, \beta)}(2x-1)=\sum_{m=0}^n c_m^{(\alpha, \beta)} x^m, \nonumber \\
&&c_m^{(\alpha, \beta)}= \frac{(-1)^{n+m}}{m!(n-m)!}
\frac{\Gamma(\beta+n+1)\Gamma(\alpha+\beta+n+m+1)}{\Gamma(\beta+m+1)\Gamma(\alpha+\beta+n+1)}
, \nonumber \\
&&(1-x)^{\alpha}= \sum_{m=0}^{\infty}
\left( \begin{array}{c} \alpha \\ m \end{array} \right)
(-x)^m,
\label{eq13}
\eeq
and we end up with the asymptotic expansion of the potential
\beq
V^{\ell_x \ell_y,\ell'_x \ell'_y}_{KL,K'L'}(\rho)\approx \delta_{LL'}\delta_{\ell_y \ell'_y}
\frac{1}{\rho^{\ell_x+\ell'_x+3}} \sum_{k=0}^{\infty} \frac{v_k}{\rho^{2k}},
\label{eq14}
\eeq
where
\beq
v_k&=&{\cal N}_K^{\ell_x \ell_y}{\cal N}_{K'}^{\ell'_x \ell_y}
\int_0^{\infty} u^{\ell_x+\ell'_x+2k+2} V\left( \frac{u}{\sqrt{\mu_{23}}} \right) du \nonumber \\
&& \times \sum_{m_1,m_2}(-1)^{k-m_1-m_2}
\left( \begin{array}{c} \ell_y+\dem \\k-m_1-m_2 \end{array} \right)
c_{m_1}^{(\ell_y+\dem,\ell_x+\dem)} c_{m_2}^{(\ell_y+\dem,\ell'_x+\dem)}.
\label{eq15}
\eeq

Owing to the finite range of the potential, the upper limit in the integrals (\ref{eq11}) has been
replaced by infinity.
Up to a normalization factor, the contribution of each $k$ value is a moment of the potential. As it is well known \cite{TDE00}, the leading term is $v_0/\rho^3$ for $\ell_x=\ell'_x=0$. Expansion (\ref{eq14}) is carried out for the three nucleus-nucleus potentials with additional Raynal-Revai transformations for the second and third terms. Analytic expansions of potentials (\ref{eq9}) are finally obtained with
\beq
\sum_{i=1}^3 V^{J\pi(Ni)}_{K  \gamma,K' \gamma'}(\rho)\approx
\frac{1}{\rho^{l_x+l'_x+3}}\sum_{k=0}^{\infty}
\frac{\tilde{v}_k}{\rho^{2k}},
\label{eq16}
\eeq
where coefficients $\tilde{v_k}$ are obtained from $v_k$ after Raynal-Revai and spin coupling transformations.

Let us evaluate coefficients $\tilde{v_k}$ for  $^6$He=$\alpha$+n+n, with the $\alpha-n$ potential taken from Kanada {\sl et al.} \cite{KKN79}.
Coefficients $\tilde{v_0}$ to $\tilde{v_4}$ are given in Table \ref{table1} for $J=0^+$.
We also provide the amplitude of the centrifugal term
\beq
v_{cent}=\frac{\hbar^2}{2m_N}(K+3/2)(K+5/2),
\eeq
which depends on $\rho$ as $1/\rho^2$. It is clear from Table \ref{table1} that coefficients $\tilde{v_k}$ are large and increasing with $k$. Integrals in (\ref{eq15}) must be computed with a high accuracy. Special attention must be paid to partial waves involving two-body forbidden states. In this case, we use a supersymmetry transform of the potential \cite{Ba87}, in order to remove forbidden states in the three-body problem. This transformation is carried out numerically, and the resulting potential presents a singularity at short distances.

From Table \ref{table1}, we evaluate the $\rho$ value where the nuclear part is negligible with respect to the centrifugal term. In other words, $\rho_{max}$ is defined as
\beq
\frac{|\tilde{v_0}|}{\rho^3_{max}}=\epsilon \times
\frac{v_{cent}}{\rho_{max}^2}.
\label{eq17}
\eeq

Values of $\rho_{max}$ are given in Table \ref{table1} by assuming $\epsilon = 0.01$. In general they are larger for low $K$ values for two reasons: $(i$) the centrifugal term is of course lower, and $(ii)$ low partial waves generally involve forbidden states which lead to singularities in the potential, and hence to larger values of $\tilde{v_0}$.

\vspace*{0.5 cm}
\begin{table}[h]
\caption{Coefficients $\tilde{v_0}$ to $\tilde{v_4}$ in $^6$He for $J=0^+,L=S=0$, and for typical partial waves (energies are expressed in MeV and lengths in fm). The bracketed values represent the power of 10,
and $\gamma=\ell_x ,\ell_y$.
\label{table1}}
\begin{tabular}{cccccccrr}
\hline
$K,\gamma$ &$K',\gamma'$ &  $\tilde{v_0}$ & $\tilde{v_1}$ & $\tilde{v_2}$ & $\tilde{v_3}$ & $\tilde{v_4}$ &$v_{cent}$&$\rho_{max}$ \\
\hline
0,0,0 & 0,0,0 &   3.40(3) & -7.46(3) & -2.02(4) & -1.53(5) & -1.78(6) &78 &4370  \\
4,0,0 & 4,0,0 &   1.18(3) & -1.20(5) & 7.31(6) & -2.13(8) & 2.87(9) & 741 &160\\
8,0,0 & 8,0,0 &   -2.59(3) & -1.19(5) & 5.46(7) & -6.66(9) & 4.98(11)& 2068 &125 \\
4,2,2 & 4,2,2 &   2.61(4) & -1.27(6) & 5.40(7) & -1.39(9) & 1.81(10)& 741 &3520 \\
8,2,2 & 8,2,2 &   5.49(4) & -7.82(6) & 1.06(9) & -1.02(11) & 6.78(12)& 2068&2660 \\
 & & & & & & & \\
0,0,0 & 4,0,0 &   -3.41(3) & 8.04(4) & -1.09(6) & 4.27(6) & 1.43(7)& &\\
0,0,0 & 8,0,0 &   1.19(3) & -1.08(5) & 6.21(6) & -1.75(8) & 2.33(9)& &\\
0,0,0 & 4,2,2 &   9.62(3) & -2.41(5) & 3.47(6) & -1.37(7) & -4.61(7)& &\\
0,0,0 & 8,2,2 &   1.40(4) & -9.90(5) & 4.80(7) & -1.30(9) & 1.71(10)& &\\
\hline
\end{tabular}
\end{table}
\vspace*{0.5 cm}

From the $\rho_{max}$ values displayed in Table \ref{table1}, it is clear that the channel radius $a$ of the $R$ matrix must be very large. Using basis functions valid up to these distances would require tremendous basis sizes. This is solved by using a propagation technique which is presented in Sec. 2.3.3.

In the analytical expansion of the potential, the maximum value $k_{max}$ is determined from the requirement
\beq
\frac{{\tilde v}_{k_{max}+1}}{a^{2k_{max}+2}}\ll
\frac{{\tilde v}_{k_{max}}}{a^{2k_{max}}}.
\label{eq36}
\eeq

This yields typical values $k_{max}\approx 3-4$, depending on the system and on the partial wave.

\subsection{Three-body $R$-matrix}
\subsubsection{Principle of the $R$ matrix}
The $R$-matrix theory is well known for many years \cite{LT58}. It allows matching a variational function over a finite interval with the correct asymptotic solutions of the Schr\"odinger equation. We summarize here the main ingredients of the $R$-matrix theory and emphasize its three-body aspects.
The $R$-matrix method is based on the assumption that the configuration space can be divided into two regions: an internal region, with radius $a$, where the solution of (\ref{eq7}) is given by some variational expansion, and an external region where the exact solutions of (\ref{eq7}) are known. This is formulated as
\beq
{\chi}^{J\pi}_{\gamma K, int}(\rho)=\sum_{i=1}^N\, c^{J\pi}_{\gamma Ki}\, u_i(\rho),
\label{eq18}
\eeq
where  the $N$ functions $u_i (\rho)$ represent the variational basis, and $c^{J\pi}_{\gamma Ki}$ are the corresponding coefficients. In the external region, it is assumed that only the Coulomb and
centrifugal potentials do not vanish; we have, for an entrance channel $\gamma' K'$,
\beq
{\chi}^{J\pi}_{\gamma K, ext}(\rho)=A^{J\pi}_{\gamma K}
\left[ H^-_{\gamma K}(k\rho)\delta_{\gamma \gamma'}\delta_{KK'}
-U^{J\pi}_{\gamma K,\gamma' K'}H^+_{\gamma K}(k\rho) \right] ,
\label{eq19}
\eeq
where the amplitude is chosen as
\beq
A^{J\pi}_{\gamma K}=i^{K+1}(2\pi/k)^{5/2},
\label{eq19b}
\eeq
and where $\ve{U}^{J\pi}$ is the collision matrix, and $k=\sqrt{2m_N E/\hbar^2}$ is the wave
number \cite{TDE00}.
If the three particles do not interact, Eq.~(\ref{eq19}) is a partial wave of a 6-dimension plane wave \cite{RR70}
\beq
\exp
\left[ i(\ve{k_x}.\ve{x}+\ve{k_y}.\ve{y}) \right] & = &
\frac{(2\pi)^3}{(k\rho)^2}  \sum_{\ell_x \ell_y LM_LK} i^K J_{K+2}(k\rho)
\nonumber \\ & & \times
{\cal Y}^{\ell_x \ell_y}_{KLM_L}(\Omega_{5\rho})\,
{\cal Y}^{\ell_x \ell_y*}_{KLM_L}(\Omega_{5k}).
\label{eq19c}
\eeq
For charged systems, we have
\beq
H^{\pm}_{\gamma K}(x)&=&G_{K+\frac{3}{2}}(\eta_{\gamma K},x) \pm iF_{K+\frac{3}{2}}(\eta_{\gamma K},x),
\label{eq20}
\eeq
where $G_{K+3/2}$ and $F_{K+3/2}$ are the irregular and regular Coulomb functions, respectively \cite{TB86}.
The Sommerfeld parameters $\eta_{\gamma K}$ are given by
\beq
\eta_{\gamma K}=z^{J\pi}_{\gamma K,\gamma K}\frac{m_N e^2}{\hbar^2 k},
\label{eq21}
\eeq
where $\ve{z}^{J\pi}$ is the effective-charge matrix (\ref{eq10}); $\eta$ therefore depends on the channel. Notice that we neglect non-diagonal terms of the Coulomb potential. This is in general a good approximation as these terms are significantly smaller than diagonal terms \cite{VNA01}.
For neutral systems, the ingoing and outgoing functions $H^{\pm}_{\gamma K} (x)$
do not depend on $\gamma$ and are defined as
\beq
H^{\pm}_{\gamma K} (x)=\pm i \left( \frac{\pi x}{2} \right) ^{1/2}
\left[ J_{K+2}(x)\pm i Y_{K+2}(x) \right] ,
\label{eq22}
\eeq
where $J_n (x)$ and $Y_n (x)$ are Bessel functions of first, and second kind, respectively.
The phase is chosen to recover the plane wave in absence of interaction (\ve{U}=\ve{I}).

For bound states ($E<0$), the external wave function is written as
\beq
{\chi}^{J\pi}_{\gamma K, ext}(\rho)=B^{J\pi}_{\gamma K}\, W_{-\eta_{\gamma K},K+2}(2\kappa\rho),
\label{eq23}
\eeq
where $W_{ab}(x)$ is a Whittaker function, and $B^{J\pi}_{\gamma K}$ the amplitude ($\kappa^2=-2m_NE/\hbar^2$). For neutral systems, we have
\beq
{\chi}^{J\pi}_{\gamma K, ext}(\rho)=C^{J\pi}_{\gamma K}\,(\kappa\rho)^{1/2} K_{K+2}(\kappa\rho),
\label{eq24}
\eeq
where $K_n (x)$ is a modified Bessel function.

\subsubsection{The Bloch-Schr\"{o}dinger equation}
The basic idea of the $R$-matrix theory is to solve Eq.~(\ref{eq7}) over the internal region. To restore the hermiticity of the kinetic energy, one solves the Bloch-Schr\"{o}dinger equation
\beq
\left( H+{\cal L}(\ve{L})-E\right) \Psi^{JM\pi} ={\cal L}(\ve{L})\Psi^{JM\pi},
\label{eq25}
\eeq
with the Bloch operator ${\cal L}(\ve{L})$ defined as
\beq
{\cal L}(\ve{L})=\frac{\hbar^2}{2m_N} \sum_{\gamma K}
|{\cal Y}^{JM}_{\gamma K}>
\delta(\rho-a_0)\frac{1}{\rho^{5/2}}\left(\frac{\partial}{\partial\rho}-\frac{L_{\gamma K}}{\rho}
\right)
\rho^{5/2}<{\cal Y}^{JM}_{\gamma K}|,
\label{eq26}
\eeq
where $\ve{L}$ is a set of arbitrary constants $L_{\gamma K}$. In the following, we assume $L_{\gamma K} = 0$ for positive energies.
Formulas presented in this subsection are given for any channel radius
$a_0$, which can be different from $a$, defined in 2.3.1.

Let us define matrix $\ve{C}^{J\pi}$ as
\beq
C_{\gamma Ki,\gamma' K'i'}^{J\pi}=<u_i {\cal Y}^{JM}_{\gamma K}
|H+{\cal L}(\ve{L})-E|
u_{i'} {\cal Y}^{JM}_{\gamma' K'}>_I ,
\label{eq28}
\eeq
where subscript $I$ means that the matrix element is evaluated in the internal region only, i.e. for $\rho \leq a_0$.
Using the partial-wave expansion (\ref{eq6}) and the continuity of the wave function at $\rho = a_0$, we
obtain the $R$-matrix at $a_0$ from
\beq
R^{J\pi}_{\gamma K,\gamma' K'}(a_0)=\frac{\hbar^2}{2m_N a_0} \sum_{i, i'}
u_i(a_0)\left( \ve{C}^{J\pi}\right)^{-1} _{\gamma Ki,\gamma' K'i'}u_{i'}(a_0).
\label{eq27}
\eeq

\subsubsection{$R$-matrix propagation and collision matrix}
As shown in Sect. 2.2, the nuclear potential extends to very large distances, even for
short-range nucleus-nucleus interactions. In other words, the asymptotic behaviour (\ref{eq19}) is not accurate below distances which may be as large as 1000 fm
or more. This is a drawback of the hyperspherical method, where even for large $\rho$ values, two particles can still be close to each other and contribute to the three-body matrix elements.

It is clear that using basis functions valid up to distances of 1000 fm is not realistic, as the size of the basis would be huge. On the other hand, using a low channel radius (typically $30 \sim 40$ fm) would keep the basis size in reasonable limits, but would not satisfy the key point of the $R$-matrix theory, namely that the wave function has reached its asymptotic behaviour at the channel radius $a_0$. This problem can be solved with propagation techniques, well known in atomic physics \cite{BN95}.
The idea is to use $a_0$ as a starting point for the $R$ matrix; its value
is small enough to allow reasonable basis sizes. The $R$ matrix is then
propagated from $a_0$ to $a$, where the Coulomb asymptotic behaviour
(\ref{eq19}) is valid. Between $a_0$ and $a$, the wave functions
$\chi^{J\pi} (\rho)$ are still given by Eq.~(\ref{eq7}), but with the potential replaced by its (analytical) asymptotic expansion.

More precisely, the internal wave functions in the different intervals are given by
\beq
{\chi}^{J\pi}_{\gamma K,int}(\rho)&=&\sum_{i=1}^N\, c^{J\pi}_{\gamma Ki}\, u_i(\rho) {\rm \ for \ } \rho\leq a_0, \nonumber \\
&=&{\tilde \chi}^{J\pi}_{\gamma K}(\rho)  {\rm \ for \ } a_0 \leq \rho \leq a,
\label{eq33}
\eeq
where ${\tilde\chi}_{\gamma K} (\rho)$ are solutions of Eq.~(\ref{eq7}) with the analytical expansion (\ref{eq16}) of the potential term.

The $R$ matrix is first computed at $a_0$ with Eq.~(\ref{eq27}) (typical values are $a_0 \approx 20-40$ fm).
Then we consider $N_0$ sets of initial conditions for ${\tilde \chi}(\rho)$, where $N_0$ is the number of $\gamma K$ values  (from now on we
drop the $J\pi$ index for clarity). We combine these sets as matrix
$\ve{{\tilde \chi}_0}(\rho)$, and choose
\beq
\ve{{\tilde \chi}_0}(a_0)=\ve{I},
\eeq
where $\ve{I}$ is the unit matrix.

According to the definition of the $R$ matrix \cite{LT58}, we immediately find the derivative at $a_0$
\beq
\ve{\tilde \chi_0}'(a_0) =\frac{1}{a_0}\ve{R}^{-1}(a_0)\ve{{\tilde \chi}_0}(a_0)=
\frac{1}{a_0}\ve{R}^{-1}(a_0).
\label{eq34}
\eeq

Knowing functions ${\tilde \chi}_{0\gamma K}$ and their derivatives at $a_0$, they are then propagated
until $a$ by using the Numerov algorithm \cite{Ra72}, well adapted to the Schr\"odinger equation. The analytical form (\ref{eq16}) of the potential is used, with a summation limited to a few $k$ values.
The $R$ matrix at $a$ is then determined by using Eq.~(\ref{eq34}) with $\ve{\tilde \chi}_{0}(a)$
and $\ve{\tilde \chi}'_{0}(a)$. We have
\beq
\ve{R}(a)=\frac{1}{a}\ve{{\tilde \chi}_0}(a)
\left( \ve{\tilde \chi_0}'(a) \right) ^{-1}.
\label{eq34b}
\eeq
Notice that the propagated $R$ matrix (\ref{eq34b}) does not depend on the choice of ${\tilde \chi}_{0}(a_0)$. In Ref.~\cite{BN95}, the propagation
is performed through the Green function defined in the intermediate region,
and expanded over a basis. The method presented here uses the Numerov algorithm, and does not rely on the choice of a basis. The analytical
form of the potential in this region makes calculations fast and accurate.

Finally the collision matrix is obtained from the $R$ matrix at the channel radius $a$ with
\beq
\ve{U}^{J\pi}=\left( \ve{Z}^{J\pi\star}\right) ^{-1} \ve{Z}^{J\pi},
\label{eq29}
\eeq
and
\beq
Z^{J\pi}_{\gamma K,\gamma' K'}=H^-_{\gamma K}(ka)\delta_{\gamma \gamma'}\delta_{KK'}
-ka (H^{-} _{\gamma K}(ka))'R^{J\pi}_{\gamma K,\gamma' K'}(a),
\label{eq30}
\eeq
where the derivation is performed with respect to $ka$.

Lower values of the channel radius $a$ can be used by employing the Gailitis method \cite{Ga76}. In this method the asymptotic forms (\ref{eq20}) are generalized with the aim of using them at shorter distances. This means that the propagation should be performed in a more limited range (typical values for $a$ are $a \sim 200-400$ fm). However this does not avoid propagation which, in any case, is very fast. In addition, the Gailitis method cannot be applied to charged systems, as it assumes from the very beginning that the coupling potentials decrease faster than $1/\rho$.

The extension of the $R$-matrix formalism to bound states is well known for two-body systems \cite{BD83}. Basically, the $L_{\gamma K}$ constants are defined so as to cancel the r.h.s. of Eq. (\ref{eq25}). Then, the problem is reduced to a matrix diagonalization with iteration on the energy \cite{BD83,DV90}.

\subsubsection{Wave functions}
Once the collision matrix is known, the internal wave function (\ref{eq33}) can be determined in both intervals. Although the choice of $\ve{{\tilde \chi}_0}(a_0)$
is arbitrary,
functions $\ve{\tilde \chi}(\rho)$ entering Eq.~(\ref{eq33}) do not depend on that
choice. In the intermediate region $a_0\leq \rho \leq a$, functions $\ve{\tilde \chi}(\rho)$
and $\ve{\tilde \chi}_0(\rho)$
are related to each other by a linear transformation
\beq
\ve{\tilde \chi}(\rho) =\ve{{\tilde \chi}_0}(\rho)\ve{M}.
\label{eq34e}
\eeq
Matrix $\ve{M}$ is deduced by using the asymptotic behaviour (\ref{eq19})
at $\rho=a$,
\beq
\ve{\tilde \chi}(a) =\ve{\tilde \chi}_0(a)\ve{M}=\ve{\chi}_{ext}(a),
\label{eq34c}
\eeq
where $\ve{\chi}_{ext}(a)$ is the matrix involving all entrance channels
[see Eq.~(\ref{eq19})]. It depends on the collision matrix.

Coefficients $c^{J\pi}_{\gamma Ki}$ defining the internal wave function in
the interval $\rho \leq a_0$ are finally obtained by
\beq
c^{J\pi}_{\gamma Ki}= \frac{\hbar^2}{2m_N}
\sum_{\gamma' K' i'}
\left( \ve{C^{-1}} \right) ^{J\pi}_{\gamma Ki,\gamma' K'i'} \,
\left( \frac{d{\tilde \chi}^{J\pi}_{\gamma' K'}}{d\rho} \right) _{\rho=a_0} \,
u_{i'}(a_0).
\eeq

\subsubsection{The Lagrange-mesh method}
Up to now, the basis functions $u_i (\rho)$ are not specified. We use here the Lagrange-mesh method which has been proved to be quite efficient in two-body \cite{HRB02} and three-body \cite{DDB03} systems. Notice however that its application to three-body continuum states is new.

When dealing with a finite interval, the $N$ basis functions $u_i (\rho)$ are defined as \cite{BHS98}
\beq
u_i (\rho)=(-1)^{N-i}\left( \frac{1-x_i}{a_0x_i} \right) ^{1/2}
\frac{\rho P_N(2\rho/a_0-1)}{\rho-a_0x_i},
\label{eq31}
\eeq
where the $x_i$ are the zeros of a shifted Legendre polynomial given by
\beq
P_N(2x_i-1)=0.
\label{eq31b}
\eeq
The basis functions satisfy the Lagrange condition
\beq
u_i(a_0x_j)=(a_0\lambda_i)^{-1/2}\delta_{ij},
\label{eq32}
\eeq
where the $\lambda_i$ are the weights of the Gauss-Legendre quadrature corresponding to the [0,1] interval, i.e. half of the weights corresponding to the traditional interval [-1,1].

The main advantage of the Lagrange-mesh technique is to strongly simplify the calculation of matrix elements (\ref{eq28}) if the Gauss approximation is used. Matrix elements of the kinetic energy $(T+{\cal L})$ are obtained analytically \cite{BHS98}. Integration over $\rho$ provides matrix elements of the potential by a single evaluation of the potential at $\rho=a_0 x_i$. The potential matrix is diagonal with respect to $i$ and $i'$.

In Ref.~\cite{DDB03}, we applied the Lagrange-mesh technique to bound states of three-body systems. As the natural interval ranges from zero to infinity, we used a Laguerre mesh. It was shown that the Gauss quadrature is quite accurate for the matrix elements, and that computer times can be strongly reduced.

\section{Applications}
\subsection{Conditions of the calculations}
Here we apply the method to the $^6$He and $^{14}$Be nuclei.
The $\alpha$-n and $^{12}$Be-n interactions are chosen as local potentials. They contain Pauli forbidden states (one in $\ell=0$ for $\alpha$-n, and one in $\ell=0,1$ for $^{12}$Be-n) which should be removed for a correct description of three-body states \cite{TDE00,DDB03}. For bound states, two methods are available: the use of a projector \cite{KP78}, and a supersymmetric transformation of the nucleus-nucleus potential \cite{Ba87}. Although both approaches provide different wave functions, spectroscopic properties are similar \cite{DDB03}. For unbound systems, it turns out that the projector technique is quite difficult to apply with a good accuracy.
Expansions similar to Eq.~(\ref{eq16}) for the projection operator provide non-local
potentials.  Consequently, all applications presented here are obtained with supersymmetric partners of the nucleus-nucleus potentials.

As collision matrices can be quite large, it is impossible to analyze all elements. To show the essential information derived from the collision matrix, we rather present some eigenphases. Those presenting the largest variation in the considered energy range are shown.

Analyzing the collision matrix in terms of eigenphases raises two problems.
First, it is in general not obvious to link the eigenphases at different energies. As eigenphases cannot be associated with given quantum numbers, there is no direct way to draw continuous eigenphases. The procedure can be strongly improved by analyzing the eigenvectors. Starting from  a given energy, eigenphases for the next energy are chosen by minimizing the differences between the corresponding eigenvectors.

A second problem associated with eigenphases arises from the Coulomb interaction. As
matrix elements of the Coulomb  force are not diagonal, the corresponding phase shifts
do not appear in a simple way, as in two-body collisions. Consequently, in order to
extract the nuclear contribution $\ve{U_N}$ from the total collision matrix $\ve{U}$, we perform
two calculations: a full calculation providing $\ve{U}$, and a calculation without the nuclear
contribution, providing the Coulomb collision matrix $\ve{U_C}$. Then we define the nuclear
collision matrix $\ve{U_N}$ by
\beq
\ve{U}=\ve{U_C}^{1/2}\ve{U_N}\ve{U_C}^{1/2}.
\label{eq37}
\eeq
As $\ve{U}$ and $\ve{U_C}$ are symmetric and unitary, the same properties hold for
$\ve{U_N}$. Examples of Coulomb phase shifts will be given in the next sections.

\subsection{Application to $^6$He and $^6$Be}
The conditions of the calculation are those of Ref. \cite{DDB03}. The $\alpha$-n potential $V_{\alpha-n}$
has been derived by Kanada {\sl et al.} \cite{KKN79}. It contains spin-orbit and parity terms. The n-n potential is the Minnesota interaction \cite{TLT77}. As bare nucleus-nucleus potentials cannot be expected to reproduce the $^6$He ground-state energy with a high accuracy, we renormalize $V_{\alpha-n}$ by a factor $\lambda = 1.051$ (note that this value was misprinted in Ref.~\cite{DDB03}).
This value reproduces the $^6$He experimental energy $-0.97$ MeV and provides 2.44 fm for the r.m.s. radius. The convergence with respect to $K_{max}$ and to the Lagrange-mesh parameters has already been discussed in Ref. \cite{DDB03}.

\begin{figure}[h]
\begin{center}
\includegraphics[width=7cm]{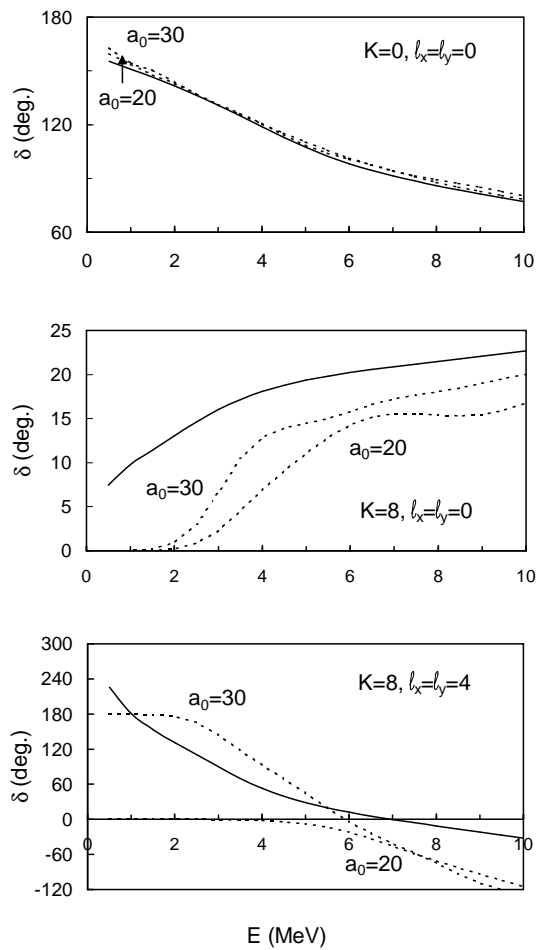}
\end{center}
\caption{$\alpha$+n+n phase shifts ($J=0^+$) for channel radii
$a_0=20$ fm $(N=20)$ and $a_0=30$ fm $(N=30)$, and for different
partial waves. Solid lines are obtained with propagation up to
$a=250$ fm of the $R$ matrix (curves corresponding to different
$a_0$ are undistinguishable), and dashed lines without
propagation. \label{fig1}}
\end{figure}


Let us first illustrate the importance of the propagation
technique. In Fig. \ref{fig1}, we plot some elements of the $J =
0^+$ collision matrix under different conditions. In each case, we
compare the phase shifts for two channel radii: $a_0 = 20$ fm and
$a_0 = 30$ fm. The calculation is performed with and without
propagation. For $K = 0$, reasonable values can be obtained
without propagation. However, for larger $K$ values ($K = 8$ is
displayed with $\ell_x =\ell_y=0$ and $\ell_x =\ell_y=4$), the
channel radius should be quite large to reach convergence. To keep
the same accuracy, the number of basis functions should be
increased. However, one basis function per fm is a good estimate,
and this leads to unrealistically large basis sizes. This
convergence problem is due to the long range of the potential. The
propagation technique (performed here up to $a=250$ fm) allows us
to get a very high stability (better than $0.1^{\circ}$ at all
energies) even for rather small channel radii. Consequently
calculations with high $K$ values are still feasible.

\begin{figure}[h]
\includegraphics[width=9cm]{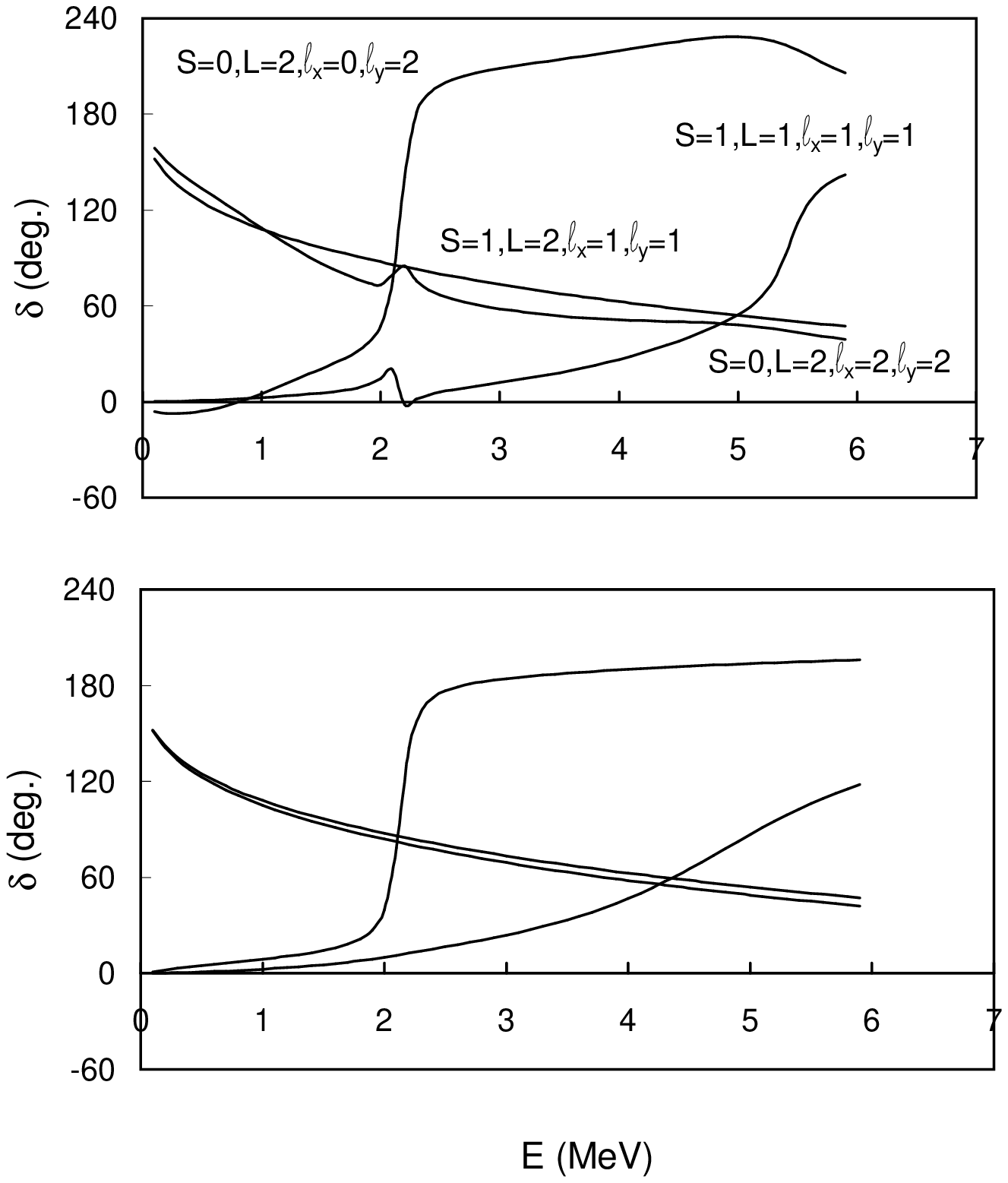}
\caption{Diagonal phase shifts (upper panel) and eigenphases
(lower panel) for the $\alpha$+n+n system ($J=2^+,K_{max}=2$).
\label{fig2}}
\end{figure}
\vspace*{0.5 cm}

To illustrate the diagonalization of the collision matrix, we compare in Fig. \ref{fig2} the diagonal phase shifts with the corresponding eigenphases. We have selected a particular case, with $J = 2^+$, and $K_{max} = 2$. With these conditions the collision matrix is $4 \times 4$, and presents a narrow resonance near 2 MeV. In the upper part of
Fig.~\ref{fig2}, we plot the diagonal phase shifts. One of them presents a $180^{\circ}$ jump, characteristical of narrow resonances. This resonant behaviour is also observable in two other partial waves. After diagonalization of the collision matrix (Fig.~\ref{fig2}, lower part) the resonant behaviour shows up in one eigenphase only. The three other eigenphases smoothly depend on energy.

\begin{figure}[h]
\includegraphics[width=9cm]{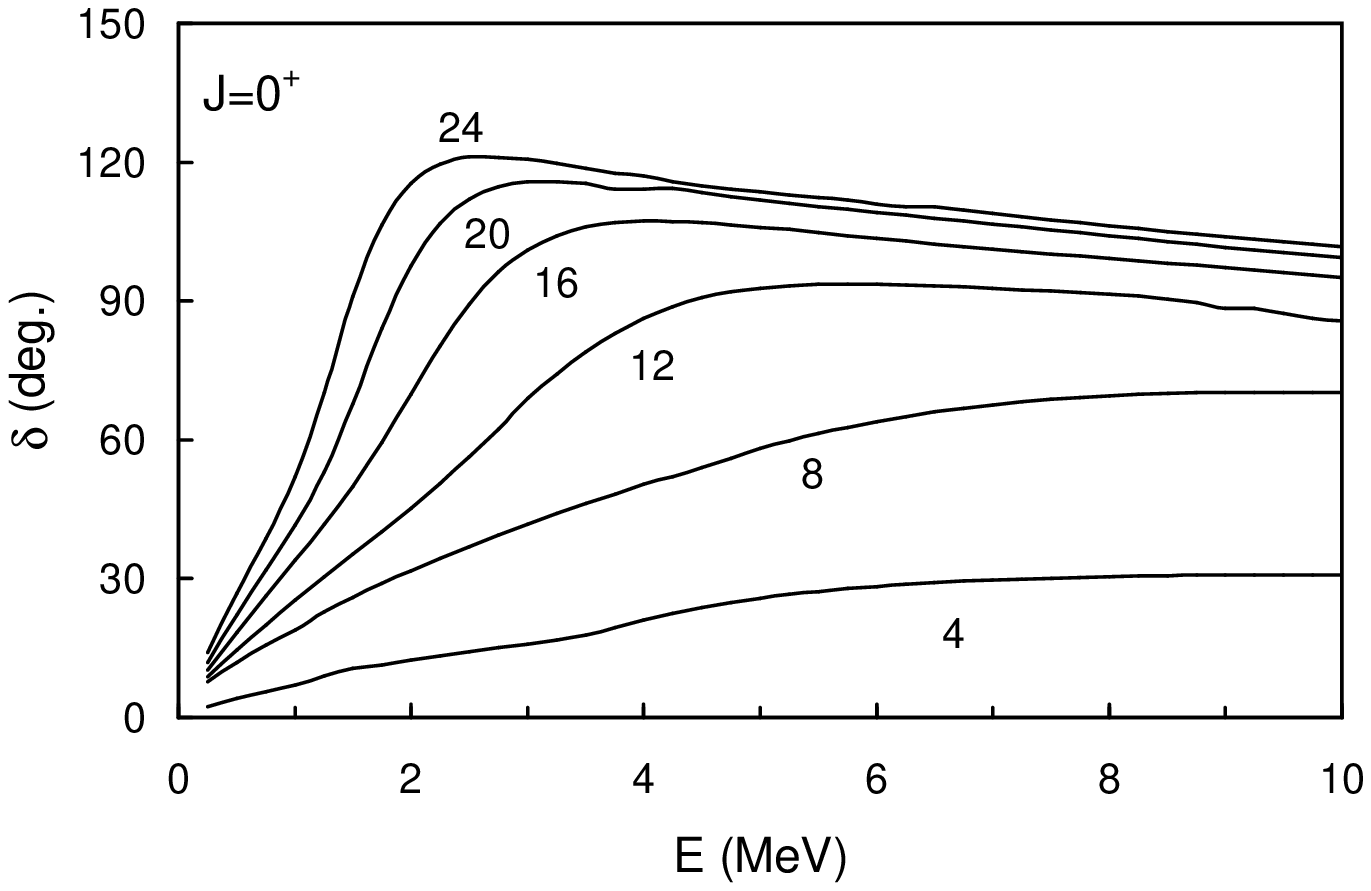}
\caption{Energy dependence of $\alpha$+n+n eigenphases ($J=0^+$)
for different $K_{max}$ values. \label{fig3}}
\end{figure}

The convergence with respect to $K_{max}$ is illustrated in Fig.~\ref{fig3} with the $J = 0^+$ eigenphases. It turns out that, at low energies, high hypermomenta are necessary to achieve a precise convergence. However, above 4 MeV, $K_{max} = 20$ is sufficient to obtain an accuracy of $2^{\circ}$.

\begin{figure}[h]
\includegraphics[width=9cm]{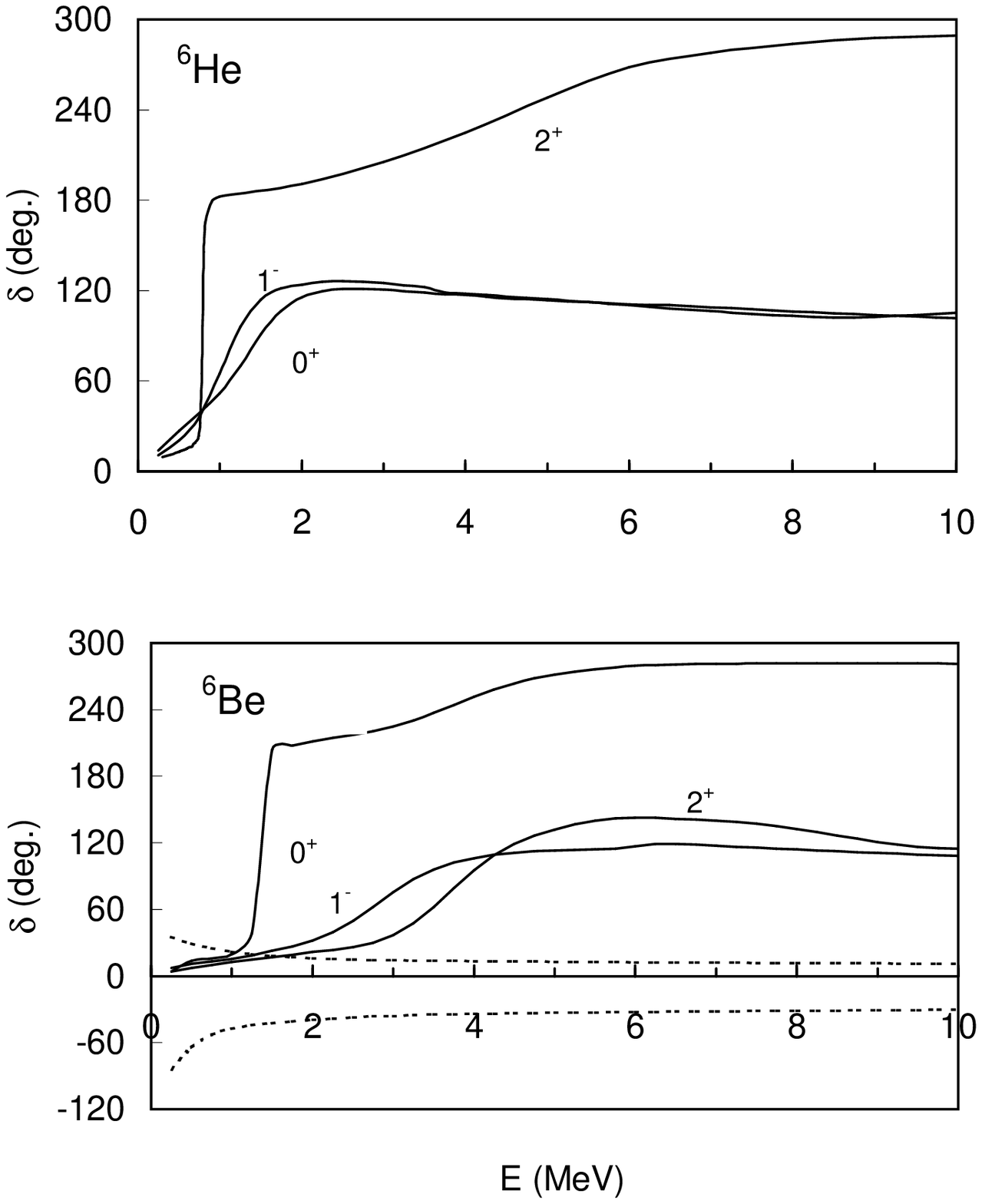}
\caption{Eigenphases of $^6$He and $^6$Be for different $J$ values (solid lines).
For $^6$Be, dotted lines represent the largest Coulomb eigenphases for $J=0^+$.
\label{fig4}}
\end{figure}

Figure \ref{fig4} gives the eigenphases for $J = 0^+, 1^-, 2^+$ in $^6$He and $^6$Be ($K_{max}$ is taken as 24, 19 and 16, respectively). As expected,
the 2$^+$ phase shift of $^6$He presents a narrow resonance. The theoretical energy
(about 0.2 MeV) is however underestimated as the experimental value \cite{TCG02} is $E = 0.82$ MeV. In order to provide meaningful properties
for this state, we have readjusted the scaling factor to $\lambda=1.020$,
which provides the correct energy.
The 0$^+$ and 1$^-$ phase shifts show broad structures near 1.5 MeV. Similar phase shifts have been obtained by Danilin {\sl et al.} \cite{DTV98,DRV04} and by Thompson {\sl et al.} \cite{TDE00}
with other potentials.

In $^6$Be, the ground state is found at $E = 1.26$ MeV with a width $\Gamma = 65$ keV.
These values are in reasonable agreement with experiment \cite{TCG02}
($E = 1.37$ MeV, $\Gamma = 92 \pm 6$ keV), the width being underestimated by the model
due to the lower energy. Experimentally, a 2$^+$ state is known near $E = 3.0$ MeV
with a width of $\Gamma = 1.16 \pm 0.06$ MeV. These properties are consistent with the theoretical 2$^+$ eigenphase, which presents a broad structure near $E \approx 4$ MeV. The largest Coulomb eigenphases ($J = 0^+$) are shown as dotted lines in Fig.~\ref{fig4}. As expected, the Coulomb interaction plays a dominant role at low energies, but it cannot be completely neglected even near 10 MeV. Coulomb eigenphases for other spin values are very similar and therefore are not presented. Energies and widths are given in Table \ref{table2}.

In Table \ref{table2}, we also present the E2 transition probability in $^6$He. For the narrow
$2^+$ resonance, we use the bound-state approximation.
Without effective charge, the $B(E2)$ value for the $0^+\rightarrow 2^+$ transition is
underestimated with respect to the experimental value \cite{AAA99}. However, the E2 matrix element is very sensitive to the effective charge. A small
correction ($\delta e=0.05 e$) provides a $B(E2)$ within the experimental error bars.

\vspace*{0.5 cm}

\vspace*{0.5 cm}
\begin{table}[h]
\caption{$^6$He and $^6$Be properties. Unless specified, experimental data are taken from Ref.~\cite{TCG02}.
\label{table2}}
\begin{tabular}{lcccc}
\hline
 & \multicolumn{2}{c}{$^6$He} & \multicolumn{2}{c}{$^6$Be}\\
\hline
  & present & exp.  & present  & exp.   \\
$E(0^+)$ (MeV) & -0.97 & -0.97 & 1.26 & 1.37 \\
$\Gamma(0^+)$ (keV)& & & 65 & $92\pm 6 $ \\
$E(2^+)$ (MeV) & 0.8 & 0.82 & $\approx 4.0$ & 3.04 \\
$\Gamma(2^+)$ (MeV) & $0.04$ & $0.113\pm 0.020 $ & $\approx 1.0$ &$1.16\pm 0.06$\\
$\sqrt{<r^2>}$ (fm)& 2.44 & $2.33\pm 0.04^{a)}$& &  \\
                   &      & $2.57\pm 0.10^{b)}$& &  \\
                   &      & $2.45\pm 0.10^{c)}$& &  \\
B(E2,$0^+ \rightarrow 2^+$) ($e^2.$fm$^4$) &1.23 ($\delta e=0$) & $3.2\pm 0.6^{d)} $ & & \\
  &2.69 ($\delta e=0.05e$) & & & \\
\hline
\end{tabular}
$^{a)}$ Ref.~\cite{THH85}, $^{b)}$ Ref.~\cite{Ch89}
$^{c)}$ Ref.~\cite{ADL04}, $^{d)}$ Ref.~\cite{AAA99}
\end{table}
\vspace*{0.5 cm}

\subsection{Application to $^{14}$Be}
As shown in previous works \cite{Ba97,ABD95,TZ96}, a $^{12}$Be+n+n three-body model can provide a realistic description of $^{14}$Be. The spectroscopy of the $^{14}$Be ground state has already been investigated in non-microscopic \cite{Ba97,ABD95,TZ96,TTT04} and microscopic \cite{De95} models. Here we extend three-body descriptions to $^{14}$Be excited and continuum states.

The $^{13}$Be ground state is expected to be a virtual $s$ wave, with a large and negative
scattering length ($a_s < -10$ fm) \cite{TYH01}. In addition, the existence of a $5/2^+$ $d$-state near 2 MeV
is well established. These properties can be reproduced by a $^{12}$Be-n potential
\beq
V(r)=-\frac{V_0+ V_s \,\ve{\ell \cdot s}}
{1+\exp((r-r_0)/a)},
\label{eq38}
\eeq
where $\ve{\ell}$ is the relative angular momentum and $\ve{s}$ the neutron spin.
In Eq.~(\ref{eq38}), $r_0 = 2.908$ fm, $a = 0.67$ fm, $V_0 = 43$ MeV, $V_s = 6$ MeV. The range and diffuseness of the Woods-Saxon potential are taken from Ref.~\cite{TZ96}. The amplitudes $V_0$ and $V_s$ provide $E(5/2^+) = 2.1$ MeV, and $a_s = -47$ fm, which are consistent with the data. For the n-n potential, we use the Minnesota interaction, as for the $^6$He study.

\begin{figure}[h]
\includegraphics[width=9cm]{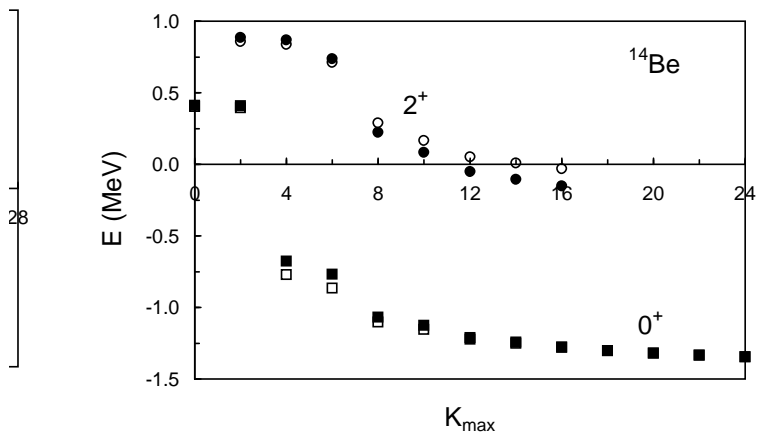}
\caption{Energy of the $^{14}$Be $0^+$ (squares) and $2^+$
(circles) states as a function of $K_{max}$. Full symbols
correspond to a renormalized $^{12}$Be-n potential, and open
symbols to a phenomenological three-body term (see text).
\label{fig5}}
\end{figure}
\vspace*{0.5 cm}

With these potentials, the $^{14}$Be ground state is found at $E = -0.16$ MeV, which represents
an underbinding with respect to experiment ($-1.34 \pm 0.11$ MeV \cite{AW93}).
This calculation has been performed with $K_{max} = 24$, which ensures the convergence.
The underbinding problem is common to all three-body approaches, and can be solved in two ways.
$(i)$ A renormalization factor $\lambda = 1.08$ provides a ground-state energy at $-1.30$ MeV,
i.e. within the experimental uncertainties. This procedure leads to a slightly bound
$^{13}$Be ground state, which might influence the $^{14}$Be properties.
$(ii)$ A three-body phenomenological term $V^{(123)}$, taken as in Ref.~\cite{TDE00},
i.e.,
\beq
V^{(123)}_{K \gamma,K' \gamma'}(\rho)=-\delta_{KK'}
\delta_{\gamma \gamma'}\, V_3/[1+(\rho/\rho_3)^2],
\eeq
reproduces the experimental energy with an amplitude $V_3 = 4.7$ MeV
(according to Ref.~\cite{TDE00}, we take $\rho_3=5$ fm).
This potential is diagonal in $(K,\gamma)$, and is simply added to the two-body term [see
Eqs.~(\ref{eq9}),(\ref{eq10})].
In $^6$He, it was shown that both readjustments of the interaction provide similar results \cite{DDB03}. However the renormalization factor is larger for $^{14}$Be, and both methods will be considered in the following.

The convergence with respect to $K_{max}$ is illustrated in Fig.~\ref{fig5}. For $J = 0^+$,
the calculations have been done with $K_{max}$ up to 24.
The energies obtained with renormalization or with
the three-body potential are very similar. This confirms the conclusion drawn for the
$^6$He nucleus \cite{DDB03}.

\vspace*{0.5 cm}
\begin{table}[h]
\caption{Properties of the $^{14}$Be $0^+$ and $2^+$ states. $\lambda$ is the renormalization
factor of the $^{12}$Be-n potential and $V_3$ is this amplitude of the three-body potential.
\label{table3}}
\begin{tabular}{llcc}
\hline
  & & $\lambda=1.08,V_3=0$ & $\lambda=1,V_3=4.7$ \\
\hline
$0^+$ &$E$ (MeV)& $-1.34$ & $-1.34$ \\
&$\sqrt{<r^2>}$ (fm)& 3.10 & 3.14 \\
&$P_{S=1}$ &0.046 & 0.033 \\
& & &\\
$2^+$ &$E$ (MeV) & $-0.15$ & $-0.03$ \\
&$\sqrt{<r^2>}$ (fm) & 2.99 & 3.04 \\
&$P_{S=1}$& 0.192 & 0.165 \\
& & & \\
$0^+ \rightarrow 2^+$&B(E2) ($e^2$.fm$^4$)&0.48 ($\delta e=0$) &0.64 ($\delta e=0$) \\
&& 3.18 ($\delta e=0.05 e$)  & 4.05 ($\delta e=0.05 e$)\\
\hline
\end{tabular}
\end{table}
\vspace*{0.5 cm}

Spectroscopic properties of $^{14}$Be are given in Tables \ref{table3} and  \ref{table4}. The r.m.s. radii have been determined with 2.57 fm as $^{12}$Be radius. For the ground state, we have $\sqrt{<r^2>} = 3.10$ fm or 3.14 fm, in nice agreement with experiment ($3.16 \pm 0.38$ fm, see Ref. \cite{TKY88}). In all cases, the $S = 1$ component (denoted as $P_{S=1}$) is small ($< 5 \%$).
The decomposition in shell-model orbitals (see Table \ref{table4}) shows that the $0^+$ state
is essentially ($\approx 70 \%$) $(2s_{1/2})^2$, with small $(2d_{3/2})^2$ and $(2d_{5/2})^2$ admixtures.

\vspace*{0.5 cm}
\begin{table}[h]
\caption{Components (in \%) in $^{14}$Be wave functions.
\label{table4}}
\begin{tabular}{lcc}
\hline
  &  $0^+(\lambda=1.08,V_3=0)$&$0^+(\lambda=1,V_3=4.7)$   \\
\hline
  $(p_{3/2})^2$ & 2.0 & 2.2    \\
  $(p_{1/2})^2$ & 1.0& 1.1    \\
  $(s_{1/2})^2$& 70.4 & 73.1     \\
  $(d_{5/2})^2$ &14.6 &13.0 \\
  $(d_{3/2})^2$ & 11.2 & 9.8    \\
 \hline
    &  $2^+(\lambda=1.08,V_3=0)$&$2^+(\lambda=1,V_3=4.7)$  \\
\hline
  $p_{3/2}p_{3/2}$ & 7.7 &8.2 \\
  $p_{3/2}f_{7/2}$ & 9.6 &9.4 \\
  $p_{1/2}p_{3/2}$ & 18.0 &18.6 \\
  $p_{1/2}f_{5/2}$ & 5.8 &5.7 \\
  $s_{1/2}d_{5/2}$ & 23.2 &23.5  \\
  $s_{1/2}d_{3/2}$ & 19.2 & 18.5  \\
  $d_{5/2}d_{5/2}$ & 5.6 &5.3 \\
  $d_{3/2}d_{5/2}$ & 4.0 &3.6  \\
  $d_{3/2}d_{3/2}$ & 3.0 &2.9\\
\hline
\end{tabular}
\end{table}
\vspace*{0.5 cm}

Regarding $J = 2^+$, we have considered values up to $K_{max} = 16$, where the number of partial waves is 172. Going beyond $K_{max} = 16$ would require too large computer memories.
Fig.~\ref{fig5} shows the energy convergence with respect to $K_{max}$.  For both potentials, the energy is below threshold, and the r.m.s. radius is close to 3 fm. A partial-wave analysis provides 19\% of $S = 1$ admixture, a value much larger than in the ground state.
Table \ref{table4} suggests that the structure of the 2$^+$ state is spread over many components. The $(s_{1/2} d_{5/2})$ component is dominant ($\approx 23 \% $) but other $(sd)$ and $(pf)$ orbitals also play a role.

E2 transition probabilities are also given in Table \ref{table3}. Without effective charge, we have B(E2,$0^+\,\rightarrow\,2^+$) = 0.48 and 0.64 $e^2$fm$^4$, which is lower than for the corresponding transition in $^6$He. However, the amplitudes of the proton and neutron E2 operators being even
more different in $^{14}$Be than in $^6$He, the B(E2) values strongly depend on the effective charge. For $\delta e = 0.05e$, we find B(E2) = 3.18 or 4.05 $e^2$fm$^4$ according to the potential.
Such transition probabilities should be measurable through Coulomb excitation experiments.

\begin{figure}[h]
\includegraphics[width=9cm]{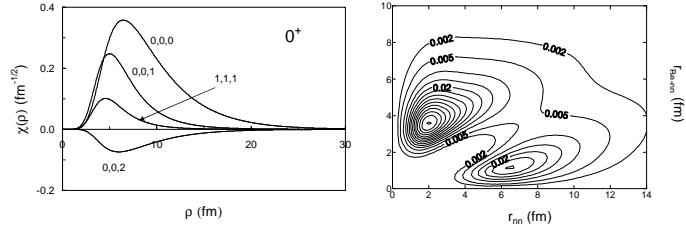}
\caption{Left panel: radial functions $\chi(\rho)$ for the $0^+$
state. The curves are labeled by $\ell_x,\ell_y,n$. Right panel:
probability $P(r_{nn},r_{Be-nn})$, deduced from Eq.~(\ref{eq39})
with $r_{nn}=\sqrt{2}x$ and $r_{Be-nn}=\sqrt{6/7}y$. Contour
levels are plotted by steps of 0.005. \label{fig6}}
\end{figure}

\begin{figure}[h]
\includegraphics[width=9cm]{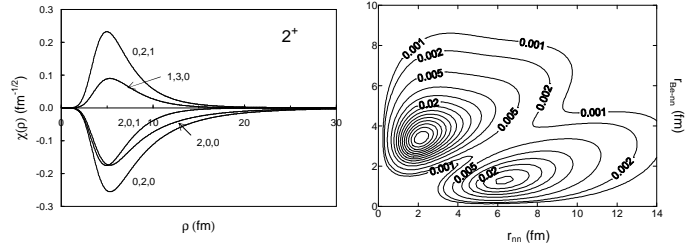}
\caption{See Fig.~\ref{fig6} for the $2^+$ state.
\label{fig7}}
\end{figure}

In Figs.~\ref{fig6}-\ref{fig7}, we present the 0$^+$ and 2$^+$ radial wave functions
and probabilities $P(x,y)$ defined as
\beq
P^{J\pi}(x,y)=\int d\Omega _x\, d\Omega _y x^2y^2\mid \Psi^{JM\pi}(\ve{x},\ve{y}) \mid ^2.
\label{eq39}
\eeq

The dominant $S = 0$ components are plotted. The $0^+$ probability shows two well distinct maxima, which resemble the maxima found in $^6$He, corresponding to "dineutron" and "cigar" configurations.
Partial waves $\chi^{J\pi}_{\gamma K}(\rho)$ have maxima for $\rho > 5$ fm.
This corresponds to distances larger than in $^6$He \cite{DDB03} where the maxima of the main components are located near 4 fm. As expected, the 2$^+$ probability is similar to the $0^+$ probability, with two maxima.

\begin{figure}[h]
\includegraphics[width=9cm]{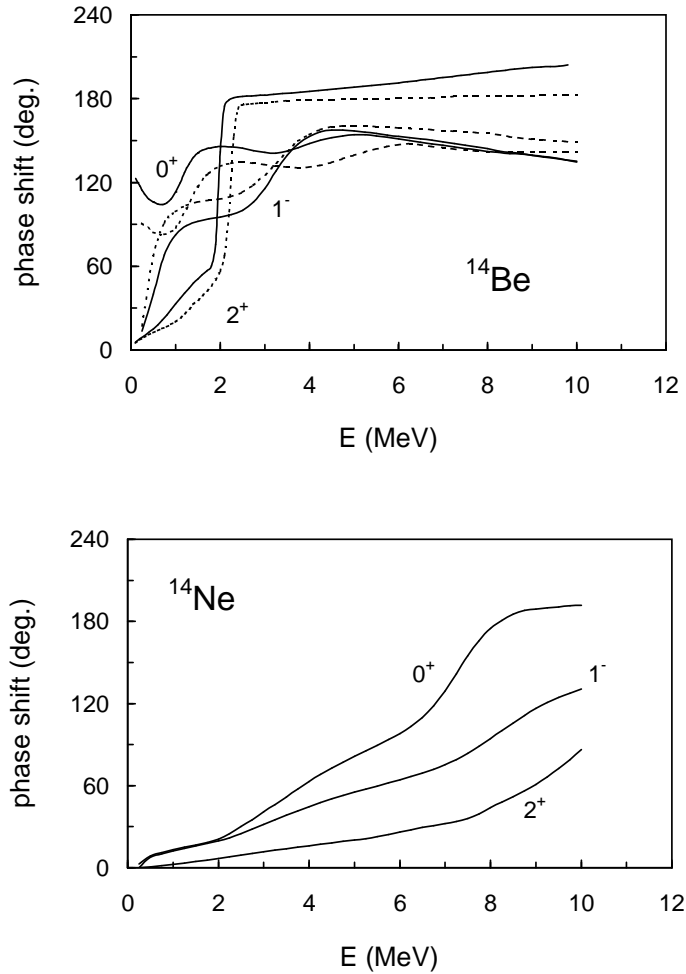}
\caption{Three-body $^{12}$Be-n-n and $^{12}$O-p-p eigenphases.
Solid lines correspond to a renormalized core-nucleon potential,
and dotted lines to a phenomenological three-body term.
\label{fig8}}
\end{figure}

Three-body eigenphases are displayed in Fig.~\ref{fig8}. As for the $^{14}$Be spectroscopy
the use of a three-body potential does not qualitatively change the phase shifts.
The 1$^-$ phase shift presents two jumps but they cannot be directly assigned to
physical resonances. On the contrary, the 2$^+$ phase shift shows a narrow resonance
near 2 MeV. For the sake of
completeness, $^{12}$O+p+p mirror phase shifts are also shown in Fig.~\ref{fig8}.
As expected, no narrow  structure is found. A very broad 0$^+$ resonance shows up near 8 MeV,
and should correspond to the $^{14}$Ne ground state.

\section{Conclusion}
In this work, we have extended the three-body formalism of Ref. \cite{DDB03} to unbound states. As for two-body systems, the Lagrange-mesh technique, associated with the $R$-matrix method, provides an efficient and accurate way to derive collision matrices and wave functions. Compared with two-body systems, three-body $R$-matrix approaches are more tedious, owing to the coupling potentials which extend to very large distances. This behaviour is inherent to the use of hyperspherical coordinates which provide three-body potentials behaving as $1/\rho^3$, even for short-range two-body interactions. This problem can be efficiently solved by using propagation techniques. Here, we propagate the wave function and the $R$ matrix by using the Numerov algorithm. This formalism has been extended to charged systems.

The $^6$He system has essentially been used as a test of the method, as most of its properties are available in the literature. The $B(E2,0^+\rightarrow 2^+)$ experimental value can be
reproduced with a small effective charge $\delta e=0.05 e$.
We have determined $\alpha$+p+p phase shifts, and found a good agreement with experiment for the $^6$Be ground-state properties.

Application to three-body $^{12}$Be+n+n states is new, and has been developed in two directions. The bound-state description of $^{14}$Be provides evidence for a 2$^+$ bound state, as expected from the shell model. The study of the $^{12}$Be+n+n system has been complemented by three-body phase shifts, which suggest the existence of a second narrow $2^+$ resonance at $E_x \approx 3.4$ MeV.

A limitation of the method is the slow convergence of the phase shifts with respect to the maximum hypermomentum $K_{max}$. To achieve a full convergence, values up to $K_{max}$ = 20 or more are necessary. This problem is even stronger for high spins, where the number of partial waves increases rapidly. A possible solution to this problem would be to apply the Feshbach reduction method \cite{Fe62} to scattering states. Another possible development would be to use a projection technique
to remove Pauli forbidden states \cite{KK77}. In that case, asymptotic potentials (\ref{eq15})
are non local, which makes the calculation still heavier.

The present model offers an efficient way to investigate bound and unbound states. In exotic nuclei, most low-lying states are unbound, and a rigorous analysis requires scattering conditions. The inclusion of the Coulomb interaction still extends the application field, and is
interesting even for non-exotic nuclei. In this context, an accurate analysis of unbound $\alpha+\alpha+\alpha$ states seems desirable in view of its strong interest in the triple-$\alpha$ reaction rate \cite{fynbo05}.

\section*{Acknowledgments}
We are grateful to Prof. F. Arickx for useful discussions about the three-body Coulomb problem.
This text presents research results of the Belgian program P5/07 on
interuniversity attraction poles initiated by the Belgian-state
Federal Services for Scientific, Technical and Cultural Affairs.
One of the authors (E.M.T.) is supported by the SSTC.


\begin{thebibliography}{}
\bibitem{Jo04} B. Jonson,  Phys. Rep. 389 (2004) 1.
\bibitem{ZDF93} M.V. Zhukov,  B.V. Danilin,  D.V. Fedorov,  J.M. Bang,  I.J. Thompson,  J.S. Vaagen,  Phys. Rep. 231 (1993) 151.
\bibitem{THH85} I. Tanihata,  H. Hamagaki,  O. Hashimoto,  Y. Shida,  N. Yoshikawa,  K. Sugimoto, O. Yamakawa,  T. Kobayashi,  N. Takahashi,  Phys. Rev. Lett. 55 (1985) 2676.
\bibitem{MF53} P.M. Morse,  H. Feshbach, Methods in Theoretical Physics,  vol. II, McGraw-Hill, New York, (1953).
\bibitem{Li95} C.D. Lin,  Phys. Rep. 257 (1995) 1.
\bibitem{DDB03} P. Descouvemont,  C. Daniel,  D. Baye,  Phys. Rev. C 67 (2003) 044309.
\bibitem{BHV02} D. Baye,  M. Hesse,  M. Vincke,  Phys. Rev. E 65 (2002) 026701.
\bibitem{AAA99} T. Aumann  {\sl et al.},  Phys. Rev. C 59 (1999) 1252.
\bibitem{Ho83} Y.K. Ho,  Phys. Rep. 99 (1983) 1.
\bibitem{KK77} V.I. Kukulin, V.M. Krasnopol'sky, J. Phys. A 10 (1977) 33.
\bibitem{LT58} A.M. Lane,  R.G. Thomas, Rev. Mod. Phys. 30 (1958) 257.
\bibitem{TDE00} I.J. Thompson,  B.V. Danilin,  V.D. Efros,  J.S. Vaagen,  J.M. Bang,  M.V. Zhukov,  Phys. Rev. C 61 (2000) 024318.
\bibitem{BN95} V.M. Burke,  C.J. Noble, Comput. Phys. Commun. 85 (1995) 471.
\bibitem{BHS98} D. Baye,  M. Hesse,  J.-M. Sparenberg,  M. Vincke,  J. Phys. B 31 (1998) 3439.
\bibitem{HSV98} M. Hesse,  J.-M. Sparenberg,  F. Van Raemdonck,  D Baye,  Nucl. Phys. A 640 (1998) 37.
\bibitem{RR70} J. Raynal, J. Revai, Nuovo. Cim. A 39 (1970) 612.
\bibitem{VNA01} V. Vasilevsky,  A.V. Nesterov,  F. Arickx,  J. Broeckhove,  Phys. Rev. C 63 (2001) 034606.
\bibitem{AS72} M. Abramowitz, I.A. Stegun, Handbook of Mathematical Functions, Dover, London (1972).
\bibitem{KKN79} H. Kanada,  T. Kaneko,  S. Nagata,  M. Nomoto,  Prog. Theor. Phys. 61 (1979) 1327.
\bibitem{Ba87} D. Baye,  Phys. Rev. Lett. 58 (1987) 2738.
\bibitem{TB86} I.J. Thompson, A.R. Barnett, J. Comput. Phys. 64 (1986) 490.
\bibitem{Ra72} J. Raynal, in "Computing as a Language of Physics", Trieste 1971, IAEA, Vienna, (1972) p. 281.
\bibitem{Ga76} M. Gailitis,  J. Phys. B9 (1976) 843.
\bibitem{BD83} D. Baye,  P. Descouvemont, Nucl. Phys. A 407 (1983) 77.
\bibitem{DV90} P. Descouvemont,  M. Vincke, Phys. Rev. A 42 (1990) 3835.
\bibitem{HRB02} M. Hesse,  J. Roland,  D. Baye,  Nucl. Phys. A 709 (2002) 184.
\bibitem{KP78} V.I. Kukulin,  V.N. Pomerantsev, Ann. Phys. 111 (1978) 330.
\bibitem{TLT77} D.R. Thompson,  M. LeMere,  Y.C. Tang,  Nucl. Phys. A 286 (1977) 53.
\bibitem{TCG02} D.R. Tilley,  C.M. Cheves,  J.L. Godwin,  G.M. Hale,  H.M. Hofmann,  J.H. Kelley,  C.G. Sheua,  H.R. Weller,  Nucl. Phys. A 708 (2002) 3.
\bibitem{DTV98} B.V. Danilin, I.J. Thompson, J.S. Vaagen, M.V. Zhukov, Nucl. Phys. A 632 (1998) 383.
\bibitem{DRV04} B.V. Danilin, T. Rogde, J.S. Vaagen, I.J. Thompson, M.V. Zhukov, Phys. Rev. C 69 (2004) 024609.
\bibitem{Ch89} L.V. Chulkov {\sl et al.}, Europhys.  Lett. 8 (1989) 245.
\bibitem{ADL04} G.D. Alkhazov, A.V. Dobrovolsky, A.A. Lobodenko, Nucl. Phys. A 734 (2004) 361.
\bibitem{Ba97} D. Baye,  Nucl. Phys. A 627 (1997) 305.
\bibitem{ABD95} A. Adahchour,  D. Baye,  P. Descouvemont,  Phys. Lett. B 356 (1995) 445.
\bibitem{TZ96} I.J. Thompson,  M.V. Zhukov, Phys. Rev. C 53 (1996) 708.
\bibitem{TTT04} T. Tarutina,  I.J. Thompson,  J.A. Tostevin,  Nucl. Phys. A 733 (2004) 53.
\bibitem{De95} P. Descouvemont,  Phys. Rev. C 52 (1995) 704.
\bibitem{TYH01} M. Thoennessen,  S. Yokoyama,  P.G. Hansen,  Phys. Rev. C 63 (2001) 014308.
\bibitem{AW93} G. Audi,  A.H. Wapstra, Nucl. Phys. A 565 (1993) 1.
\bibitem{TKY88} I. Tanihata, T. Kobayashi, O. Yamakawa, S. Shimoura, K. Ekuni, K. Sugimoto, N. Takahashi, T. Shimoda, H. Sato,  Phys. Lett. B 206 (1988) 592.
\bibitem{Fe62} H. Feshbach, Ann. Phys. 19 (1962) 287.
\bibitem{fynbo05} H. Fynbo {\sl et al.}, Nature 433 (2005) 136.
\end{thebibliography}
\end{document}